\documentclass[aip,pof,reprint]{revtex4-1}
\usepackage{bm,graphicx,amsmath,times}

\newcommand{\bk}{\boldsymbol{k}}
\newcommand{\bx}{\boldsymbol{x}}
\newcommand{\bnabla}{\boldsymbol{\nabla}}
\newcommand{\bcdot}{\boldsymbol{\cdot}}

\newcommand{\dd}{{\rm d}}
\newcommand{\ee}{\mathrm{e}}
\newcommand{\ii}{\mathrm{i}}
\newcommand{\Ku}{\mathrm{Ku}}
\newcommand{\OW}{\mathrm{OW}}
\newcommand{\OWth}{\OW_{\mathrm{th}}}

\newcommand{\zext}{\zeta^\dagger}

\newcommand{\zgvr}{\zeta_{\mathrm{GVR}}}

\newcommand{\DD}{\mathcal{D_\mathrm{ls}}}

\newcommand{\E}[1]{Eq.~(\ref{#1})}
\newcommand{\F}[1]{Fig.~\ref{#1}}
\newcommand{\FF}[1]{Figure~\ref{#1}}

\newcommand{\T}[1]{TABLE~\ref{#1}}

\newcommand{\R}[1]{Ref.~\onlinecite{#1}}

\newcommand{\avg}[1]{\left\langle#1\right\rangle}
\newcommand{\bs}[1]{\boldsymbol{#1}}

\DeclareMathOperator{\sgn}{sgn}
 
\begin{document}

\title{Non-universal velocity probability densities in two-dimensional turbulence: the effect of large-scale dissipation}

\author{Yue-Kin Tsang}

\affiliation{Scripps Institution of Oceanography, University of California, San Diego, La Jolla, California, 92093 USA}

\date{\today}

\begin{abstract}
We show that some statistical properties of forced two-dimensional turbulence have an important sensitivity to the form of large-scale dissipation which is required to damp the inverse cascade. We consider three models of large-scale dissipation: linear ``Ekman" drag, non-linear quadratic drag, and scale selective hypo-drag that damps only low-wavenumber modes. In all cases, the statistically steady vorticity field is dominated by almost axisymmetric vortices, and the probability density function of vorticity is non-Gaussian. However, in the case of linear and quadratic drag, we find that the velocity statistics is close to Gaussian, with non-negligible contribution coming from the background turbulent flow. On the other hand, with hypo-drag, the probability density function of velocity is non-Gaussian and is predominantly determined by the properties of the vortices. With hypo-drag, the relative positions of the vortices and the exponential distribution of the vortex extremum are important factors responsible for the non-Gaussian velocity statistics.
\end{abstract}

\maketitle

\section{Introduction}

The probability density function (PDF) of velocity components is an important characterization of the statistics of turbulence. Numerical studies of two-dimensional turbulence by Bracco {\it et~al.}\cite{Bracco00} and Pasquero {\it et~al.}\cite{Pasquero01} have found strongly non-Gaussian velocity PDFs in the decaying and forced cases respectively. Recently, Bandi {\it et~al.}\cite{Bandi09} and Tsang and Young\cite{Tsang09} reported the contrary observation of Gaussian velocity PDFs in simulations of two-dimensional turbulence forced at the small scales. Experiments of electromagnetically driven two-dimensional turbulence by Jun {\it et~al.}\cite{Jun06} also show that the velocity PDFs are close to Gaussian, albeit with slightly sub-Gaussian tails. One of the goals of the present work is to provide a possible explanation to the different results found in the literature.

The larger issue here is the identification of universal features of forced, statistically steady two-dimensional turbulence. A fully developed inverse cascade requires the removal of energy at large length-scales. There are different processes, with varying degrees of physical justification, which might dissipate energy at large scales so that a statistical steady state can be achieved. 

In sections~\ref{simulation} and \ref{statistics} we describe numerical simulations of two-dimensional turbulence equilibrated by three different large-scale dissipative mechanisms:
\begin{enumerate}
\item[(a)]  linear drag, also known as Rayleigh or Ekman friction;
\item[(b)] ``quadratic drag'', produced by  three-dimensional boundary-layer turbulence;
\item[(c)] ``scale-selective'' or hypo-drag, acting only on low-wavenumber modes.
\end{enumerate}
In section \ref{statistics}, we show that in all three cases the statistically steady vorticity field is dominated by long-lived almost axisymmetric vortices. The  signature of these coherent structures is a non-Gaussian tail on the vorticity PDF. However the non-Gaussian tail in case (c) is longer and stronger than in the other two cases, and we show that only in this case are velocity statistics strongly non-Gaussian. This result implies that one-point velocity statistics is not a universal property of forced two-dimensional turbulence and one must take into account the large-scale dissipation mechanism employed when interpreting such statistics.

In section~\ref{role}, we investigate the role of vortices in determining the different velocity statistics observed in the simulations presented here. We first extract the properties of the vortices by means of a vortex census algorithm, and then decompose the vorticity field into vortical and background components. We find that only with hypo-drag do the vortices have dominant influence on the shape of the velocity PDF. For linear and quadratic drag, the background vorticity plays an essential role in determining the velocity PDF. Section~\ref{conclusion} is the conclusion.

\section{A model of forced two-dimensional turbulence}
\label{simulation}

\begin{table*}[t]
\caption{A summary of the statistical properties of nine runs. All quantities are non-dimensionalized using $k_{\!f}^{-1}$ to scale length and $\tau_{\! f}$ to scale time. In the third column $\varepsilon_{\mathrm{ls}}$ is the rate of energy dissipation by large-scale drag (as opposed to hyperviscosity), and in the fourth column $E \equiv \avg{u^2+v^2}/2$. }
\vskip4mm
\centering
\begin{tabular}{cccccccccccc}
\hline\hline
  & $\varepsilon = \avg{v f}$ & $\varepsilon_{\mathrm{ls}}/\varepsilon$ & $E$ & $\sqrt{\avg{u^2}}$ & $\sqrt{\avg{v^2}}$ & $\sqrt{\avg{\zeta^2}}$ & $\Ku_{u}$ & $\Ku_v$ & $\Ku_\zeta$ & $t_0$ & $T$ \\
\hline
$\alpha$=0.003 & 0.1333 & 0.9434 & 35.21 & 5.928 & 5.940 & 2.812 & 6.230 & 6.409 & 66.56 & 3000 & 5500\\
$\alpha$=0.007 & 0.1623 & 0.9442 & 26.96 & 5.171 & 5.212 & 3.142 & 7.258 & 7.490 & 71.90 & 2000 & 6500\\
$\alpha$=0.015 & 0.1862 & 0.9448 & 22.16 & 4.686 & 4.730 & 3.262 & 7.880 & 7.824 & 58.12 & 500 & 4000\\
\hline
$\mu$=0.003 & 0.1400 & 0.9504 & 22.17 & 4.712 & 4.703 & 2.144 & 3.252 & 3.266 & 17.95 & 4000 & 4000\\
$\mu$=0.007 & 0.1894 & 0.9531 & 12.89 & 3.584 & 3.594 & 2.214 & 3.057 & 3.104 & 11.73 & 4000 & 8000\\
$\mu$=0.015 & 0.2544 & 0.9607 & 8.145 & 2.852 & 2.843 & 2.282 & 2.955 & 3.045 & 8.343 & 600 & 2400 \\ 
\hline
$\kappa$=0.001 & 0.1822 & 0.9528 & 13.00 & 3.598 & 3.613 & 2.135 & 2.878 & 2.945 & 10.18 & 200 & 4000\\
$\kappa$=0.003 & 0.2531 & 0.9613 & 7.865 & 2.782 & 2.824 & 2.221 & 2.812 & 2.900 & 7.422 & 500 & 4000\\
$\kappa$=0.007 & 0.3382 & 0.9709 & 5.457 & 2.295 & 2.374 & 2.269 & 2.794 & 2.899 & 5.975 & 200 & 4000\\
\hline\hline
\end{tabular}
\label{table}
\end{table*}

We consider a two-dimensional incompressible flow $(u,v)$ driven by a single-scale unidirectional steady body force, $\bs f(x)=\tau_f^{-2}k_f^{-1}\sin(k_fx)\hat{\bm y}$. The velocity is given in terms of a streamfunction $\psi$ as $(u,v)=(-\psi_y,\psi_x)$ and the vorticity is $\zeta=\nabla^2\psi$. The forced vorticity equation is therefore
\begin{equation}
\zeta_t +u \zeta_x + v \zeta_y = \tau_{\!f}^{-2} \cos k_{\!f} x  - \DD -  \nu  \nabla^{8} \zeta  \, .
\label{qg1}
\end{equation}
Large-scale drag $\DD$ is the main dissipative mechanism for the energy supplied by $\bs f$;  the hyperviscosity $\nu$ removes enstrophy at the small scales.\cite{Frisch08} The domain is a doubly periodic square $2\pi L\times 2\pi L$. In pursuit of an inverse cascade driven by small scale forcing $\bs f$,  we take  $L=32/k_{\!f}$ so that there is a reasonable separation between the forcing scale and the domain scale.

To achieve a statistically steady equilibrium, energy must be  removed  at large scales by $\DD$. It is instructive to consider different models of $\DD$. A natural and simple choice, which applies to experiments such as those summarized in \R{Tabeling02},  is linear drag
\begin{equation}
\DD = \mu \zeta\, ;
\label{linearDrag}
\end{equation}
$\mu$ has dimensions inverse time. Another possibility, with geophysical motivation discussed in \R{Grianik04}, is quadratic drag
\begin{equation}
\DD = \kappa \bnabla \!\bcdot \!\left( |\bnabla \psi | \bnabla \psi \right)\, ;
\label{nonlinearDrag}
\end{equation}
$\kappa$ has dimensions inverse length.

The drag in Eqs.~(\ref{linearDrag}) and (\ref{nonlinearDrag}) acts on all modes and therefore results in some degree of leakage from the Kraichnan-Batchelor inverse cascade. To achieve an idealized realization of the inverse cascade, some authors have used a spectral prescription in which drag is applied only to modes with small wavenumber.\cite{Maltrud93} Thus if vorticity is represented as 
\begin{equation}
\zeta = \sum_{\bk} \tilde \zeta_{\bk} \ee^{\ii \bk \bcdot \bx}
\end{equation}
then, in spectral space, this ``hypo-drag" is  
\begin{equation}
\widetilde{\DD} =
\begin{cases}
\alpha \tilde \zeta_{\bk} \, , & \text{if $|\bk| \leq k_{\alpha} $,}\\
0 \, ,& \text{otherwise;}
\end{cases}
\label{draggy}
\end{equation}
$\alpha$ above has dimensions inverse time. We use $k_{\alpha} = 3k_f/16$ in our implementation of hypo-drag. 

Another means of selectively damping the low-wavenumber modes is hypo-viscosity $\DD = - \eta \psi$ used in \R{Pasquero01}. We have verified that simulation results using hypo-viscosity (not shown) are similar to those with hypo-drag, and thus restrict attention to $\DD$ in \E{draggy} as  representative of scale-selective drag.

\begin{figure}[t] 
\begin{center}
\includegraphics[width=8.3cm]{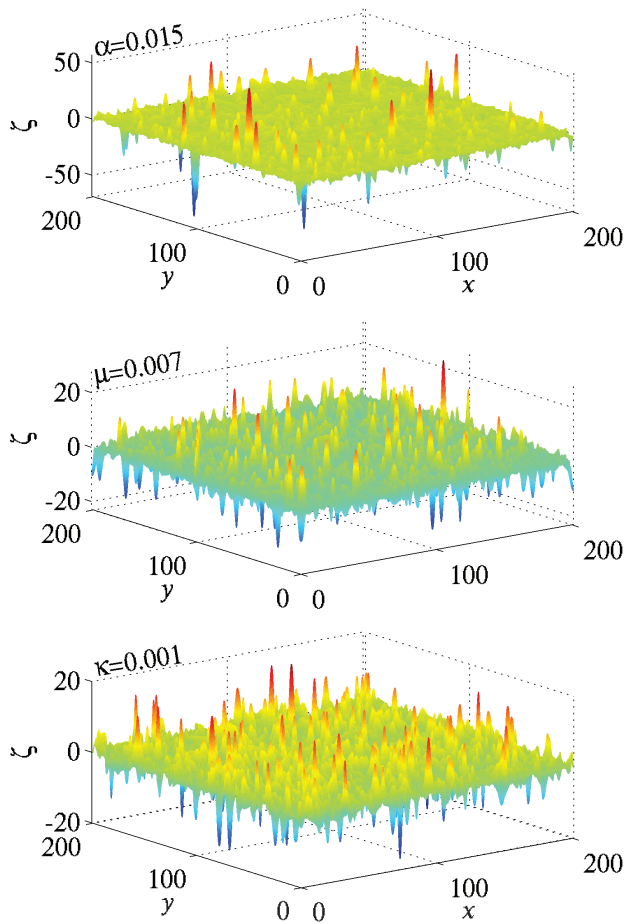}
\end{center}
\caption{(Color online) Snapshots of the vorticity from simulations using different models of $\DD$. Upper panel  is hypo-drag with $\alpha=0.015$; middle panel is linear drag with $\mu=0.007$; lower panel is quadratic drag with $\kappa=0.001$. Strong vortices are indicated by the larger scale used on the $\zeta$-axis in the top panel.}
\label{vorty}
\end{figure}

We have solved \E{qg1} using the pseudo-spectral method with exponential time differencing.\cite{Cox02,Kassam05} All simulations have $\nu=10^{-5}$, $L=32/k_f$, and resolution  $1024^2$. \FF{vorty} shows snapshots of the vorticity field using the  different models of $\DD$ described above.

\section{Statistics of the simulations}
\label{statistics}

Statistically steady turbulent solutions are obtained after an initial transient period $t_0$.\cite{Tsang09} The most important statistic characterizing forced-dissipative turbulence is the average energy injection rate $\varepsilon$, defined as:
\begin{equation}
\varepsilon \equiv \avg{v f} \equiv  \frac{1}{(2 \pi L)^2T} \int_{t_0}^{t_0+T} \!\!\!\! \iint \!\! v f \, \dd x \dd y  \dd t\, .
\label{epsilon}
\end{equation}
\T{table} shows the values of $\varepsilon$  for some representative runs. Other statistical quantities reported in \T{table} are space-time averaged $\avg{\cdot}$ using the recipe above, and all PDFs presented in this article are obtained by analyzing about 80 snapshots taken in the time interval between $t_0$ and $t_0+T$.

There are three runs in \T{table}, with different models of $\DD$, that all happen to have $0.18 < k_f^{2}\tau_f^{3} \varepsilon < 0.19 $. Because these runs have roughly the same energy injection, one can regard them as equivalently forced. \FF{vorty} shows snapshots of the vorticity field in these three cases, and indicates that the flow is populated by coherent vortices no matter which model is used for $\DD$. An extensive literature characterizes both forced and decaying two-dimensional turbulence in terms of the mutual advection and merger of vortices. This phenomenology applies to the solutions in \F{vorty}: close encounters of like-signed vortices result in merger, and new vortices are created by nucleation out of the sinusoidal pattern of vorticity impressed by the $\cos k_{\! f} x$ forcing. 

\begin{figure}[t] 
\begin{center}
\includegraphics[width=8.3cm]{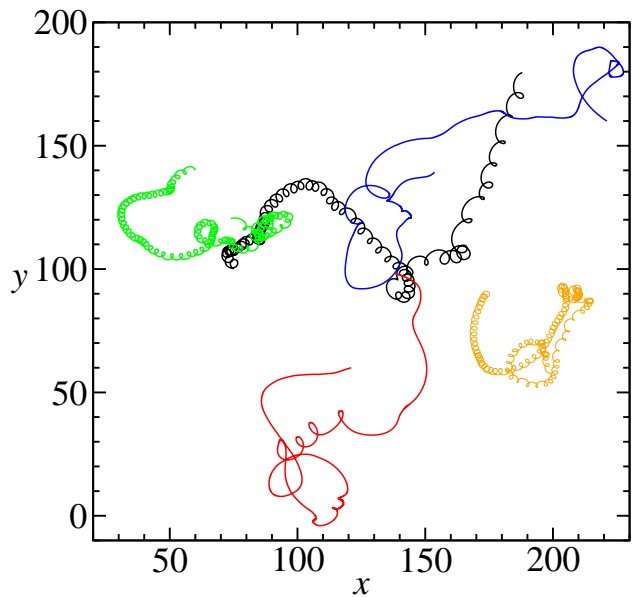}
\end{center}
\caption{(Color online) Lagrangian trajectories of selected particles from the simulation  shown in the middle panel of \F{vorty} (linear drag $\mu$). Three of the particles were released close to a vorticity extremum and exhibit  ``looping'' trajectories, characteristic of trapping within a vortex. The other two particles were released in  the filamentary sea between the vortices. }
\label{traj}
\end{figure}

In \F{vorty} the presence of strong vortices in the top panel is indicated by the scale used on the $\zeta$-axis, which is larger by over a factor of two than the scale in the lower two panels: with hypo-drag, the vortices stand above the background of incoherent vorticity more clearly than with linear and quadratic drag. Thus it is reassuring to demonstrate that the maxima in the lower panels of \F{vorty} are indeed long-lived coherent structures. A defining characteristic of a coherent vortex is that the structure survives for many turnover times, and during this lifetime the structure traps and transports fluid particles over appreciable distances. The Lagrangian trajectories plotted in \F{traj} shows that the $\zeta$-extrema in the middle panel of \F{vorty}, with $\DD = \mu \zeta$, satisfy this criterion, trajectories in runs using the other two models of $\DD$ are very similar (not shown). Thus, although the vorticity extrema in the lower two panels of \F{vorty} are smaller than those in the top panel, Lagrangian trajectories confirm the coherent nature of the vortices in all three flows.

\FF{ek} shows energy spectra corresponding to the three runs in \F{vorty}. In all three cases there is a rough $k^{-5/3}$ range, though this range is not more developed in the hypo-drag case. In all cases there is a strong inverse cascade, as evinced by the third column of \T{table} which shows that $\DD$ is responsible for dissipating over $94\%$ of the injected energy. The most remarkable result in \F{ek} is the coincidence of the spectra corresponding to linear and quadratic drag. This anticipates a main conclusion of this work: simulations with quadratic and linear drag have very similar statistical properties, while hypo-drag differs in many respects from these physically-based dissipative mechanisms.

We now turn to the PDFs of vorticity and velocity. The top panel of \F{pdfz1} shows vorticity PDFs and provides a quantitative characterization of the different vortex strengths evident in \F{vorty}: all three PDFs have non-Gaussian tails, but these tails are much stronger in the case of hypo-drag. In \T{table} the kurtosis of the vorticity field,
\begin{equation}
\Ku_\zeta \equiv {\avg{\zeta^4}}/{\avg{\zeta^2}^2}\, ,
\label{vkosis}
\end{equation}
conveniently summarizes the degree of non-Gaussianity of the three cases. With hypo-drag $\Ku_\zeta> 55 \gg 3$; runs with quadratic drag have the smallest vorticity kurtosis. Nonetheless, $\Ku_\zeta$ is significantly larger than the Gaussian value $3$ in all cases.

\begin{figure}[t] 
\begin{center}
\includegraphics[width=8.3cm]{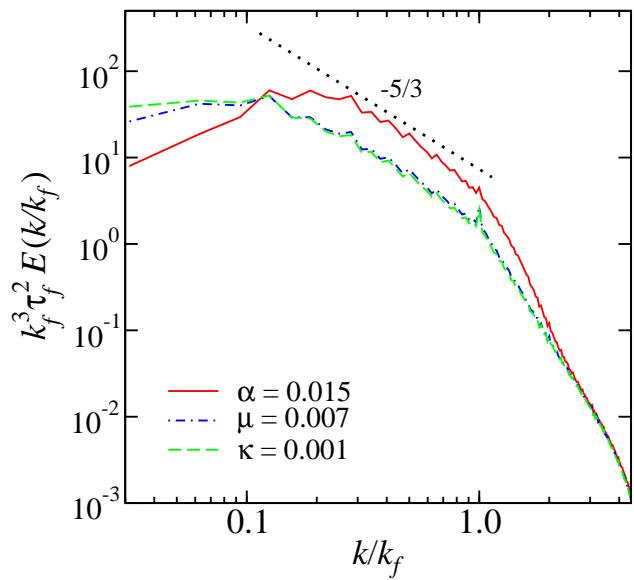}
\end{center}
\vspace{-0.3cm}
\caption{(Color online) Energy spectra for the the three simulations shown in \F{vorty}.}
\label{ek}
\end{figure}

\begin{figure}[t] 
\begin{center}
\includegraphics[width=8.3cm]{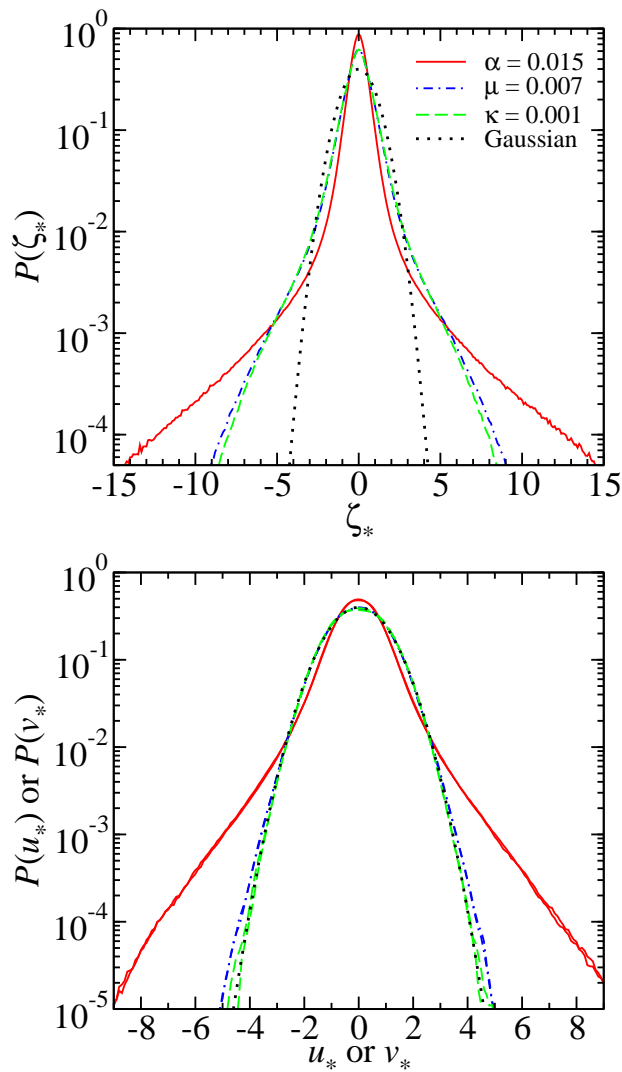}
\end{center}
\caption{(Color online) Top panel: vorticity PDFs corresponding to the three simulations in \F{vorty}. The abscissa is $\zeta_* \equiv \zeta/\sqrt{\avg{\zeta^2}}$, and the dotted parabola is a standard Gaussian. Lower panel: velocity PDFs, and the standard Gaussian (the dotted parabola). The abscissa is normalized with $u_*\equiv u/\sqrt{\avg{u^2}}$ and $v_*\equiv v/\sqrt{\avg{v^2}}$. PDFs of $u$ and $v$ almost coincide, and are close to the standard Gaussian for linear drag and quadratic drag.}
\label{pdfz1}
\end{figure}

Velocity PDFs are shown in the lower panel of \F{pdfz1}. Notice that the PDF of $u$ is almost indistinguishable from the PDF of $v$ --- in \T{table} $\avg{u^2}$ and $\avg{v^2}$ typically differ by less than $1\%$ --- indicating that the flows in \F{vorty} are almost isotropic, despite the anisotropic $\cos k_{\!f} x$ forcing in \E{qg1}.

The lower panel of \F{pdfz1} shows the most striking difference between hypo-drag and the other two cases: with hypo-drag, the velocity PDF is strongly non-Gaussian. The velocity kurtosis, $\Ku_u$ or $\Ku_v$, in \T{table} is greater than $6$ in the case of hypo-drag, and close to $3$ in the cases of linear and quadratic drag. Thus the PDF of velocity is not a universal statistical feature of forced-dissipative two-dimensional turbulence.

\FF{pdfuv} shows how the PDFs of vorticity and velocity vary in a sequence of simulations using linear drag $\mu \zeta$, as the non-dimensional drag coefficient $\mu \tau_{\! f}$ is varied by a factor of 50. With $\mu \tau_{\! f}=0.1$, both the velocity and the vorticity PDF are close to the standard Gaussian. It is known\cite{Tsang08} that the linear stability threshold of the steady laminar solution of \E{qg1} is $\mu \tau_{\!f} = 0.52$. Thus even the run with $\mu \tau_{\!f}=0.1$ in \F{pdfuv} is strongly unstable. As $\mu \tau_{\! f}$ decreases the non-Gaussian tails in the vorticity PDF become stronger, but there is little alteration of the velocity PDF, which always remains relatively close to the standard Gaussian. The trend in the upper panel of \F{pdfuv} indicates that the vortices become stronger as the drag is reduced, yet this has only a little impact on the velocity statistics in the lower panel. In particular, there is no evidence that the statistical properties of this sequence of simulations with linear drag resemble those of the hypo-drag in the limit $\mu \tau_{\! f} \to 0$.

\begin{figure}[t] 
\begin{center}
\includegraphics[width=8.3cm]{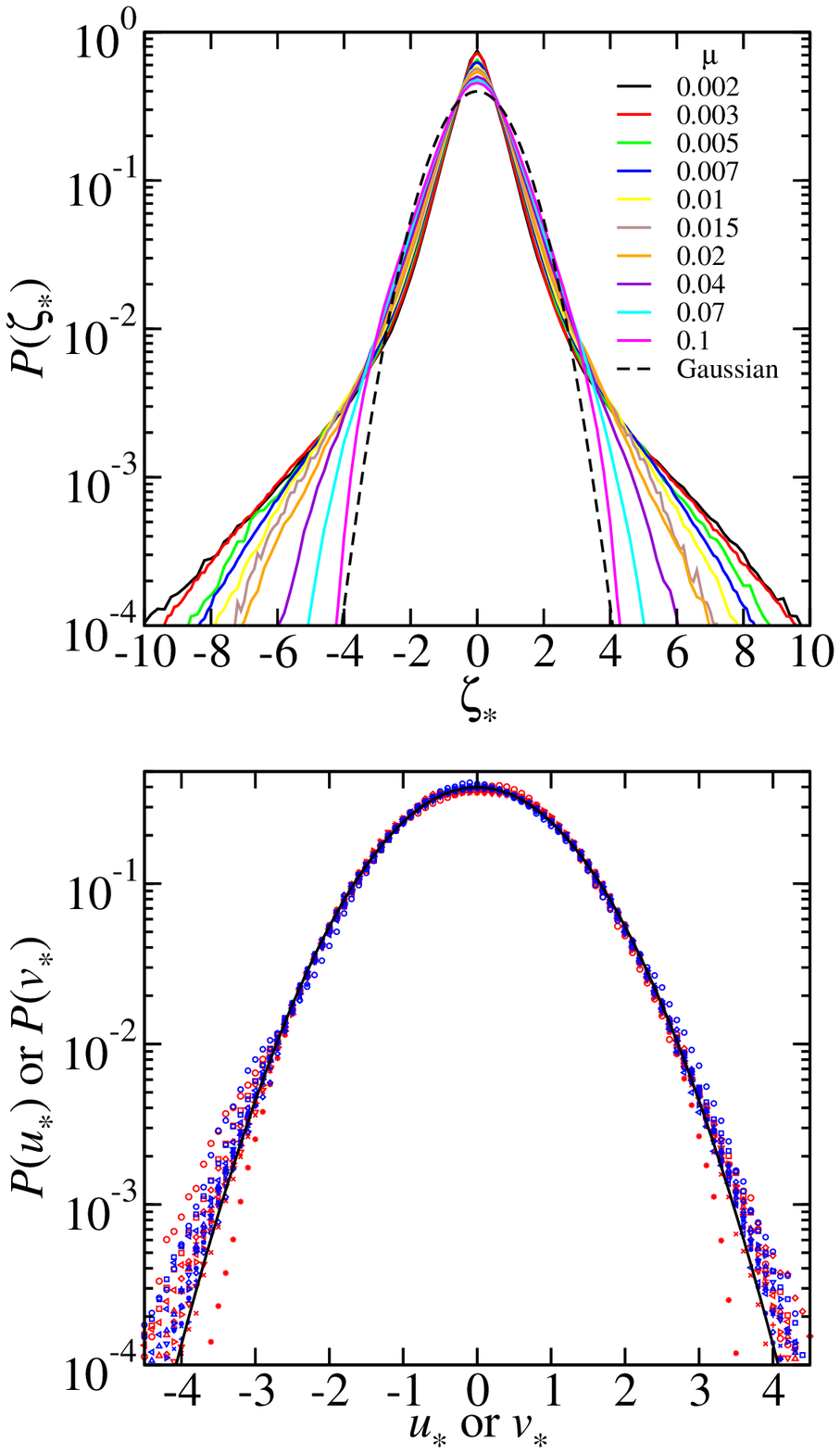}
\end{center}
\caption{(Color online) Vorticity (upper panel) and velocity (lower panel) PDFs from a sequence of simulations using linear drag, with $\mu \tau_{\!f}$ decreasing from $0.1$ to $0.002$. The abscissa uses standardized variables $u_* \equiv u/\sqrt{\avg{u^2}}$ etc. The velocity PDF remains close to the standard Gaussian (the smooth parabola) at all values of $\mu \tau_{\!f}$, while the vorticity PDF becomes increasingly non-Gaussian as $\mu \tau_{\!f}$ is decreased.}
\label{pdfuv}
\end{figure}

The main point of this descriptive section is that in the lower panel of \F{pdfz1}, the velocity PDFs corresponding to linear and quadratic drag are close to Gaussian, while the hypo-drag velocity PDF has strong non-Gaussian tails. These tails are even evident in the PDF of hypo-drag \textit{streamfunction} (not shown). In other words, the vortices in the upper panel of \F{vorty} are so extreme that they are expressed as isolated axisymmetric extrema in snapshots of $\psi$. By contrast, the streamfunctions corresponding to the lower two panels of \F{vorty} resemble diffuse clouds resulting from aggregations of like signed vortices.\cite{Tabeling02}

\section{Role of coherent vortices in determining the velocity PDF}
\label{role}

\begin{figure}[t] 
\begin{center}
\includegraphics[width=8.3cm]{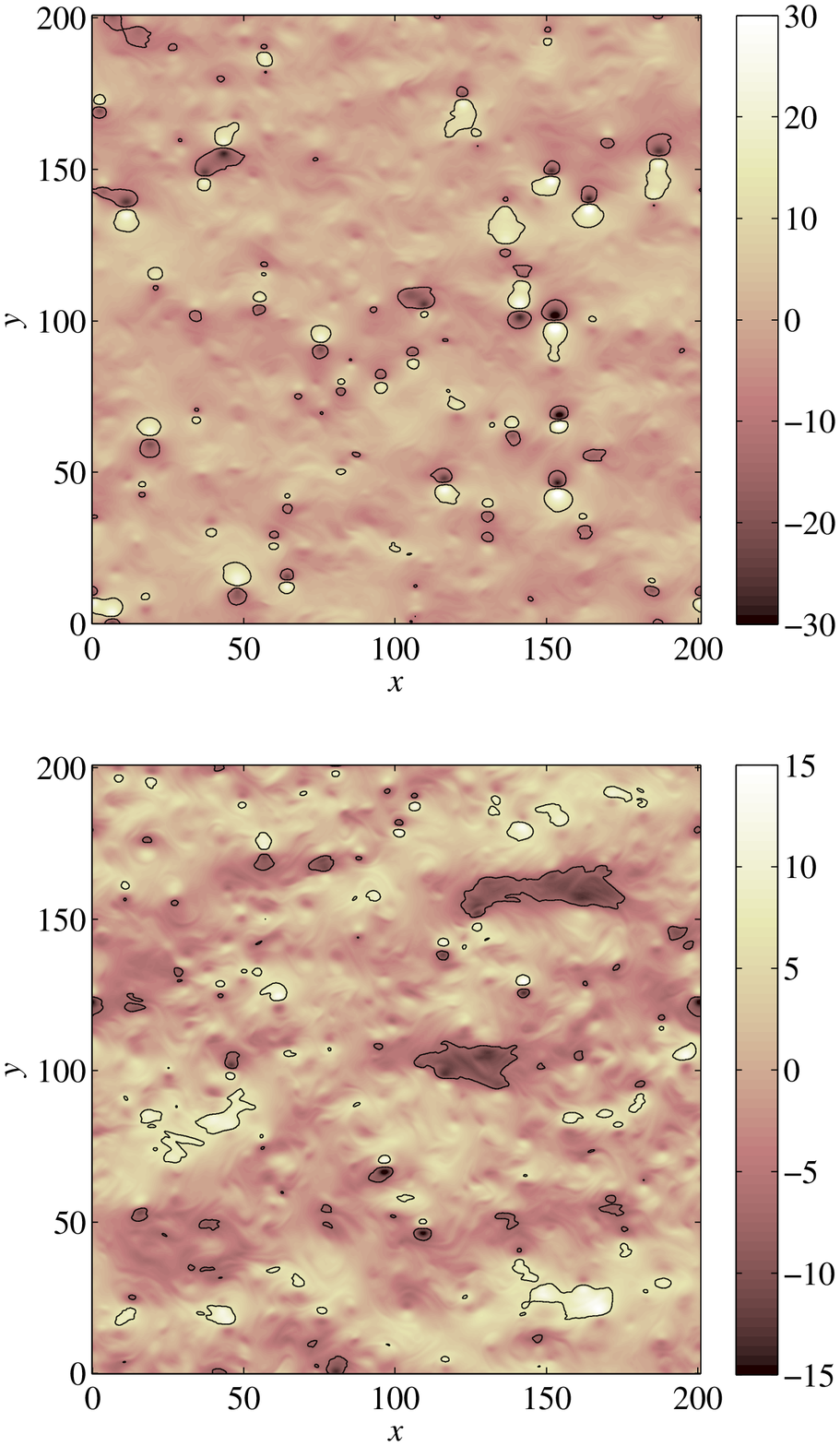}
\end{center}
\caption{(Color online) Snapshots of the velocity $u$ for hypo-drag $\alpha = 0.015$ (upper) and linear drag $\mu=0.007$ (lower); regions in which $|u| \geq 2 \sqrt{\avg{u^2}}$ are inside the black solid bubbles.}
\label{u_condu}
\vspace{0.3cm}
\end{figure}

In this section, we try to give some insights into how large-scale dissipation $\DD$ affects the velocity PDF. The first clue comes from snapshots of the velocity field $u(x,y,t)$ such as those shown in \F{u_condu}. We have enclosed regions in which $|u|$ exceeds the value $2\sqrt{\avg{u^2}}$ by black solid bubbles. It is clear that with hypo-drag $\alpha$ most of these high-velocity regions are adjacent to the vortices and resemble the velocity profile generated by either an isolated vortex or by a few interacting vortices. Hence, with hypo-drag, the tails of the velocity PDF are due to the dominant influence of the vortices and their interaction.

With linear drag $\mu$, the high-velocity regions are often not associated with individual vortices but instead occupy large amorphous space between vortices. It appears that the velocity field generated by the vortices is sometimes masked by the strong background strain flow. This suggests that in this case, both the vortices and the strain field contribute to the Gaussian velocity PDF observed.

\subsection{Vortex census and Gaussian vortex reconstruction}

To provide further evidence for the picture described above, we estimate the contribution to the velocity PDF from the vortices. A common way to do this is to partition in physical space the vorticity field into non-overlapping sets of vortical and non-vortical components according to whether the vorticity or the Okubo-Weiss parameter (defined in \E{ow} below) exceeds a certain threshold.\cite{Bracco00,Petersen06} We find that such simple partitioning does not work well for the forced-dissipative turbulence studied here, especially for the case with linear or quadratic drag. It is because for each individual vortex, the vorticity drops from a high value at its center to roughly the background level on its boundary. For the flows considered here, the non-vortical components include short-lived filaments or irregular patches with vorticity above the background level. This makes it difficult to find a threshold that can cleanly isolate the vortices. Therefore we employ a different approach described as follow.

We conducted an automated census of the vortex population and determined the following properties of the flow: the number of vortices $N$, the location $(x_i,y_i)$, the vorticity extremum $\zext_i$ and an appropriately defined radius $a_i$ of each vortex. Following previous works,\cite{Petersen06,Thompson06} the vortex census algorithm used here  is based on thresholds in the Okubo-Weiss parameter,\cite{Okubo70,Weiss91} defined as
\begin{equation}
\OW \equiv (u_x-v_y)^2 + (u_y+v_x)^2 - \zeta^2 \,.
\label{ow}
\end{equation}
$\OW$ is negative where rotation dominates and positive where strain dominates. Details of the implementation and further results from our algorithm will be reported elsewhere. Briefly, we identify vortex centers as local vorticity extrema that have $\OW/\sigma_{\OW}$ less than a certain threshold $\OWth$, where $\sigma_{\OW}$ is the standard deviation of $\OW$ over all space and time. The radius $a_i$ of a vortex is determined based on the distance from the vortex center to the closest contour with $\OW=0$ (hence the factor 1.1209 in \E{z_gvr} below). We only include structures that are roughly circular, so a single radius $a_i$ is representative of the  size of vortex $i$. 

With the census data, we then construct a synthetic vorticity field $\zgvr$ as:
\begin{equation}
\zgvr(x,y,t) = \sum_{i=1}^{N}\zext_i\exp\!\left[-\frac{(x-x_i)^2+(y-y_i)^2}{(a_i/1.1209)^2}\right] + C_0
\label{z_gvr}
\end{equation}
where $C_0(t)$ is a small constant to ensure the spatial average of $\zgvr$ vanishes for all $t$. We call this {\it Gaussian vortex reconstruction} (GVR). \E{z_gvr} represents the contribution of the vortices to the total vorticity field. Inspecting the profiles of the vortices in the simulations convinces us that the Gaussian model used in \E{z_gvr} is a good approximation for most vortices. With $\zgvr$ in \E{z_gvr}, the velocity PDF can then be computed and compared with that of the simulation. 

\begin{figure}[t] 
\begin{center}
\includegraphics[width=8.3cm]{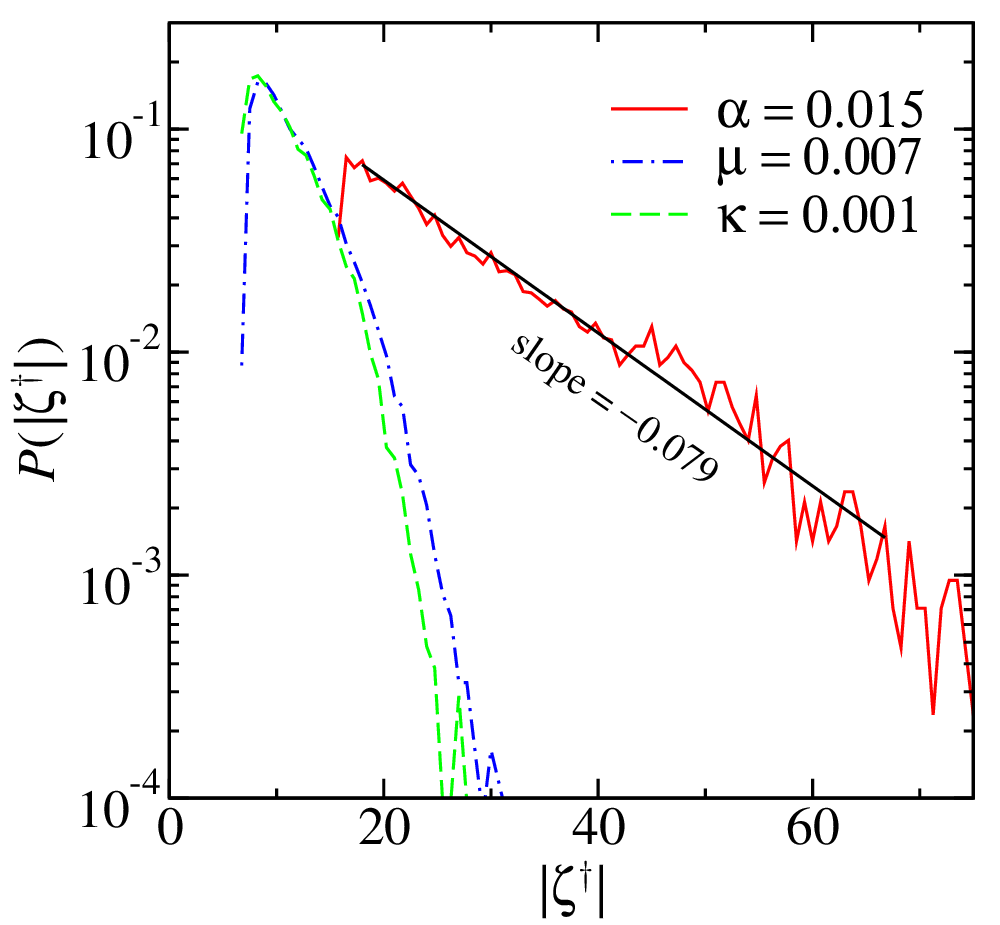}
\end{center}
\caption{(Color online) Distribution of vorticity extremum $|\zext|$ for hypo-drag $\alpha$, linear drag $\mu$ and quadratic drag $\kappa$ simulations. The solid straight line is a least-square-fit to the exponential.}
\label{pdf_zext}
\end{figure}

The vortex census reveals a striking difference in the distribution of vorticity extremum $P(|\zext|)$ between the hypo-drag case and the other two cases. As shown in \F{pdf_zext}, $P(|\zext|)$ has a long exponential tail for the hypo-drag case. In contrast, with linear and quadratic drag, the distribution of $|\zext|$ is much narrower and has approximately Gaussian tails. The distribution $P(|\zext|)$ will turn out to be important in determining the shape of the velocity PDF. Also notice the near coincidence of $P(|\zext|)$ for linear and quadratic drag, showing once again that the two systems have similar statistical properties. The curves in \F{pdf_zext} are obtained using $\OWth=3$. The threshold $\OWth$ sets a lower bound on $|\zext|$ for which a point will be counted as a vortex center. Hence changing $\OWth$ will only change the lower cutoff to the distribution $P(|\zext|)$ in \F{pdf_zext}, but not its shape for larger values of $|\zext|$.

\subsection{Vortical contribution to velocity PDF with hypo-drag}

We now estimate the velocity PDF induced by the vortices in our simulations by computing the velocity PDF using the GVR vorticity \E{z_gvr}. The doubly periodic boundary condition is implemented in solving this Poisson equation. Although the number of vortices identified by the census, $N$, is sensitive to $\OWth$, we verified that for all three models of drag used here, the velocity PDF, and especially its tail, is largely independent of $\OWth$ over a wide range of $\OWth$. \FF{pdfu_ow} demonstrates such independence in the case of hypo-drag. For the results presented below, we use the threshold $\OWth=3$, knowing this subjective choice does not strongly affect velocity statistics obtained from $\zgvr$.

\begin{figure}[t] 
\begin{center}
\includegraphics[width=8.3cm]{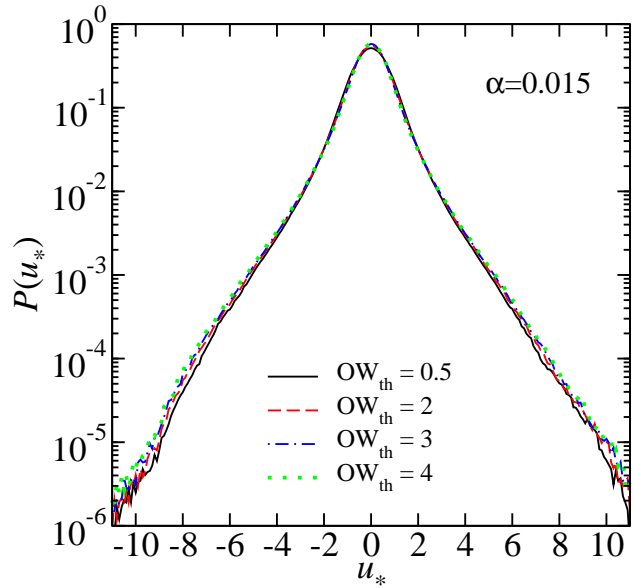}
\end{center}
\caption{(Color online) Insensitivity of velocity PDF obtained by the Gaussian vortex reconstruction \E{z_gvr} to the threshold $\OWth$ for hypo-drag $\alpha$. Results are similar for linear and quadratic drag.}
\label{pdfu_ow}
\end{figure}

The upper panel of \F{pdfu_gvr} shows the results for hypo-drag. The field $\zgvr$ produces a velocity PDF that is almost identical to the one induced by the full vorticity in numerical simulation. This proves that with hypo-drag the velocity PDF is predominately controlled by the vortices.

\begin{figure}[t] 
\begin{center}
\includegraphics[width=8.3cm]{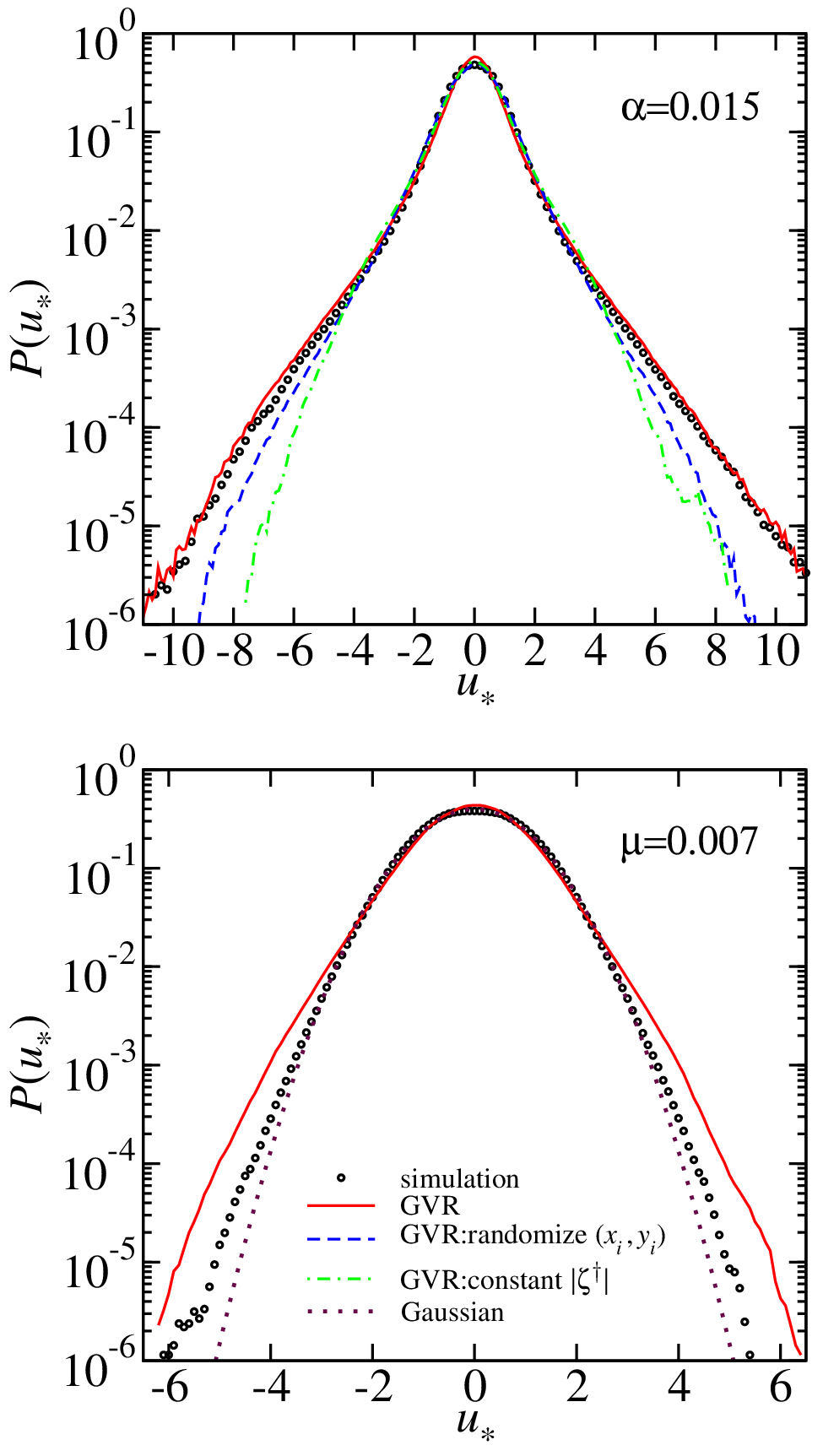}
\end{center}
\caption{(Color online) Comparison of velocity PDF from simulations (circles) to that obtained by the Gaussian vortex reconstruction \E{z_gvr} (solid) for hypo-drag (upper) and linear drag (lower). Velocity PDFs obtained by GVR but with randomized vortex positions $(x_i,y_i)$ (dashed) and the exponential distribution in vorticity extremum $P(|\zext|)$ replaced by constant $|\zext|$ (dot-dashed) are also plotted in the hypo-drag case.}
\label{pdfu_gvr}
\end{figure}

Further in the upper panel of \F{pdfu_gvr}, the GVR velocity PDF does not agree with simulation if we replace $\zext$ with $\sgn(\zext)$ in \E{z_gvr} so that all vortices have the same constant $|\zext|$. Here $\sgn(\cdot)$ is the signum function that extracts the sign of its argument. This shows that the structure of the exponential density of $|\zext|$ in \F{pdf_zext} is crucial in determining the shape of the non-Gaussian tail of the velocity PDF. 
 
If one randomizes the positions $(x_i,y_i)$ of the vortices in $\zgvr$, then again in the upper panel of \F{pdfu_gvr}, the GVR velocity tails do not agree with simulation. This indicates that correlations between the positions of the vortices are also essential in determining tail velocity statistics.
We resisted this conclusion because it indicates that vortex-gas models,\cite{Min96,Jimenez96,Chavanis00} which assume no correlations between vortex positions, cannot explain the observed non-Gaussian velocity tails. Extensive comparisons between different Monte Carlo vortex models (of which the GVR is the most complete) and simulations finally drove us to accept that uncorrelated vortex-gas models cannot explain the observed velocity tails. 

Further insight can be gained by inspecting the PDF $P(\delta r)$ of vortex pair separation $\delta r$. We define $\delta r$ as the distance between the two vortex extrema of a vortex pair. We compute $P(\delta r)$ separately for opposite-signed vortex pairs and like-signed vortex pairs. For an ensemble of vortices with no correlation in position, $P(\delta r)$ is linear in $\delta r$ for $\delta r/(2\pi L) < 0.5$. \FF{pdfvortsep} clearly shows that {\it close} vortex pairs (small $\delta r$) in hypo-drag flows are more probable to have opposite signs and less probable to have the same signs as compared to a vortex-gas ensemble with no correlations between vortex positions. It is likely that the presence of vortex dipoles and vortex mergers are  the source of the observed position correlations.  Dynamics of close vortex dipoles has also been studied in a point-vortex model from the Lagrangian viewpoint.\cite{Weiss98}

\begin{figure}[t] 
\begin{center}
\includegraphics[width=8.3cm]{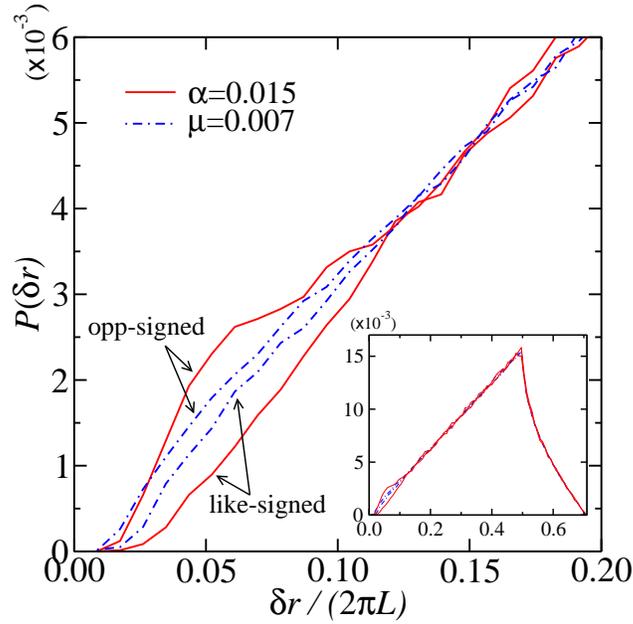}
\end{center}
\caption{(Color online) Probability distribution $P(\delta r)$ of vortex pair separation $\delta r$ for hypo-drag $\alpha$ and linear drag $\mu$. Opposite-signed and like-signed vortex pairs are considered separately. The main figure focuses on small $\delta r$ and the inset shows $P(\delta r)$ for the full range of $\delta r$.}
\label{pdfvortsep}
\end{figure}

\subsection{Vortical contribution to velocity PDF with linear drag}

For linear drag, the lower panel of \F{pdfu_gvr} shows that the GVR velocity PDF is non-Gaussian and thus does not agree with the velocity PDF computed from the simulation data. This implies that with linear drag, the background turbulent flow outside the vortices plays a non-negligible role in shaping the velocity PDF. In \F{pdfvortsep}, we plot the PDF of vortex pair separation, $P(\delta r)$.
At small $\delta r$, $P(\delta r)$ for opposite-signed vortex pairs and that for like-signed pairs are relatively close to each other, both being approximately linear in $\delta r$. Hence, we see that with linear drag, the vortices evolve roughly independent of each other and they do not have a dominant effect on the velocity PDF. This is in sharp contrast to the hypo-drag case. The results for quadratic drag (not shown) are similar to those with linear drag.

Experience with the GVR velocity PDF demonstrates the different roles played by the vortices depending on the form of large-scale dissipation. \FF{pdfu_gvr} confirms the qualitative description based on \F{u_condu} given at the beginning of this section.

\subsection{Transition from hypo-drag to linear drag}

\begin{figure}[t] 
\begin{center}
\includegraphics[width=8.3cm]{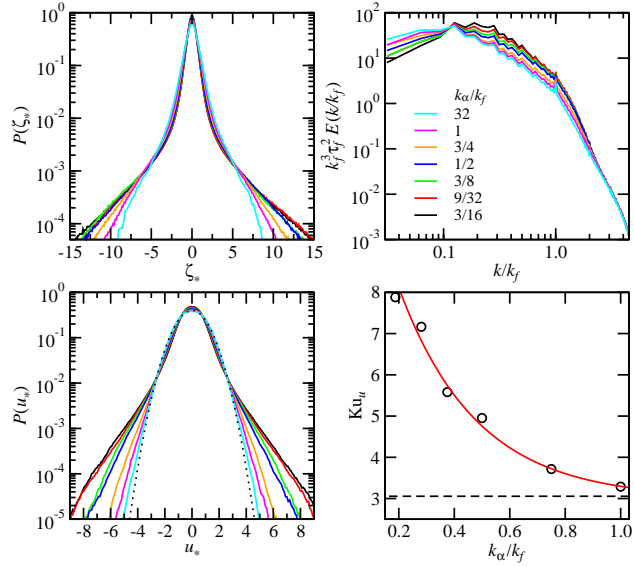}
\end{center}
\caption{(Color online) The evolution of standardized vorticity PDF $P(\zeta_*)$, standardized velocity PDF $P(u_*)$, energy spectrum $E(k)$ and velocity kurtosis $\Ku_{u}$ when the value of $k_\alpha$ in hypo-drag simulations is varied. $k_\alpha=32 k_f$ corresponds to the simulation with linear drag $\mu=0.007$. The dotted curve in the lower left panel is the standard Gaussian. In the lower right panel, the dashed line indicates the value of $\Ku_{u}$ for linear drag $\mu=0.007$; the solid curve is the fit in  \E{Ku_fit}.}
\label{k_alpha}
\end{figure}

Linear drag damps all modes equally, while  hypo-drag acts only on modes with $|\bs k| < k_\alpha$, where $k_\alpha$ is the cut-off in Eq. \eqref{draggy}. It is interesting to see how the statistics change as $k_\alpha$ systematically increases from the standard value $k_\alpha = 3 k_f/16$ used previously in this work. As we increase $k_\alpha$, we keep the energy injection rate $\varepsilon$ at approximately $0.18$  by adjusting the coefficient $\alpha$ within the range $0.007 \leq\alpha\leq 0.015$. \FF{k_alpha} summarizes the results for seven different $k_\alpha$. The maximum wavenumber in our numerical representation is $32 k_f$ and thus  $k_\alpha=32 k_f$ is equivalent to uniform linear drag with $\mu=0.007$. However there are negligible changes in statistics once $k_\alpha>k_f$ i.e., hypo-drag acting on  $k\leq k_f$  is effectively  uniform.

In \F{k_alpha} we see that the vorticity PDF $P(\zeta_*)$ and the velocity PDF $P(u_*)$ evolve gradually as $k_\alpha$ increases toward $k_f$. As expected, $P(\zeta_*)$ has shorter tails for larger $k_\alpha$, indicating that the number of extremely strong vortices decreases as more modes are  damped. Accordingly, $P(u_*)$ changes from non-Gaussian to almost-Gaussian as $k_\alpha$ increases. Comparatively, the changes in both $P(\zeta_*)$ and $P(u_*)$ are minimal for $k_f < k_\alpha < 32 k_f$. The lower right panel in \F{k_alpha} shows the velocity kurtosis $\Ku_u$ versus $k_\alpha$. $\Ku_u$ drops like an exponential initially and saturates at about $3$ once $k_\alpha > k_f$. The  empirical fit
\begin{equation}
\Ku_u \approx 3 + 10.40\, \ee^{-3.56(k_\alpha/k_f)} \,.
\label{Ku_fit}
\end{equation}
summarizes the simulation results. We note that while the shape of the vorticity and velocity PDFs changes with $k_\alpha$, there is no drastic difference in the general shape of the energy spectrum $E(k)$ for different $k_\alpha$. For all  $k_\alpha$ considered here, $E(k)$ exhibits a similar scaling range with slope approximately equals $-5/3$.

\section{conclusion}
\label{conclusion}

The numerical simulations presented here show that some basic statistical properties of two-dimensional turbulence have an important sensitivity to the form of large-scale dissipation. In particular, we find that the velocity PDF is not a universal feature of forced-dissipative two-dimensional turbulence in the inverse cascade regime. All three types of dissipative mechanisms studied here result in vorticity fields populated by vortices and produce a non-Gaussian vorticity PDF. However the hypo-drag velocity statistics are strongly non-Gaussian while for linear and quadratic drag, the velocity PDF is close to Gaussian. Thus the different velocity statistics reported in Refs.~\onlinecite{Bracco00,Pasquero01,Bandi09,Tsang09} are explained by the different dissipative mechanisms employed by these authors: Tsang and Young\cite{Tsang09} and Bandi {\it et~al.}\cite{Bandi09} use linear drag, while Pasquero {\it et~al.}\cite{Pasquero01} employ scale-selective drag. Bracco {\it et~al.}\cite{Bracco00} study freely evolving two-dimensional turbulence, with no forcing and no large-scale drag at all. Previous studies on two-dimensional turbulence with linear drag\cite{Danilov01,Boffetta02,Tsang05} have also found non-universal features depending on the drag coefficient.

It is interesting that the vortices play very different roles in the velocity statistics depending on the form of the drag. The vortices in the hypo-drag case are much stronger and have dominant effects on the velocity PDF. The relative positions of the vortices and the exponential distribution in the vorticity extremum are important in controlling the shape of the non-Gaussian velocity PDF. In contrast, for flows using linear or quadratic drag, the background turbulent flow is essential in determining the velocity statistics. This means that the Gaussian velocity PDFs observed in these cases are not consequences of the application of the central limit theorem to the vortices. One can also conclude from the above results that vortex-gas models\cite{Min96,Jimenez96,Chavanis00} that do not take into account position correlations, resulting from vortex interactions, or the contribution from the background flow, cannot explain the velocity PDFs reported here.

The strong vortices observed in forced flows with hypo-drag closely resemble the vortices that characterize freely-evolving two-dimensional turbulence.\cite{Carnevale91,Bracco00} In both situations the velocity statistics are strongly non-Gaussian. However this cannot be used as an argument for hypo-drag in specific applications. For electromagnetically driven flows in shallow fluid layers,\cite{Jun06} such as those reviewed by Tabeling,\cite{Tabeling02} linear drag is a strong dissipative mechanism.\cite{Burgess99} For oceanographic applications, some form of large-scale dissipation is necessary to stop the observed inverse cascade,\cite{Scott05} and there is no geophysical justification for hypo-drag. In a study of baroclinic turbulence,\cite{Arbic04} it is shown that bottom Ekman drag must be adjusted to a moderate value to produce eddy statistics that match observed ocean flows. Thompson and Young\cite{Thompson06,Thompson07} demonstrate the importance of Ekman drag in determining the baroclinic eddy heat flux. Sen {\it et~al.}\cite{Sen08} have recently argued that turbulent bottom boundary layer drag (quadratic drag) is a significant sink of ocean kinetic energy. These results argue in favor of linear and quadratic drag in oceanographic applications. Interestingly, for the statistical properties reported here, we find little difference between linear and quadratic drag, in agreement with earlier studies.\cite{Grianik04,Arbic08}

\begin{acknowledgments}
I acknowledge useful discussions and correspondence with Antonello Provenzale and Jost von Hardenberg. I thank Bill Young for comments and discussions that improve the scientific content and presentation of the paper. This work was supported by the National Science Foundation under Grant No.~OCE07-26320.
\end{acknowledgments}

\bibliography{velpdf}

\end{document}